\documentclass[10pt]{olplainarticle}
\usepackage{textgreek,graphicx}
\usepackage{rotating}
\usepackage{float}
\usepackage{xr}
\usepackage{hyperref}
\usepackage[normalem]{ulem}
\usepackage{stmaryrd}
\expandafter\def\csname opt@stmaryrd.sty\endcsname
{only,shortleftarrow,shortrightarrow}
\usepackage{extpfeil}
\usepackage{ctable} 

\definecolor{mycolor}{RGB}{0, 0, 0}
\definecolor{REFcolor}{rgb}{0, 0, 0}
\usepackage{moreverb} 

\title{An interaction network approach predicts protein cage architectures in bionanotechnology}

\author[1,2,*]{Farzad Fatehi}
\author[1,2,3,*]{Reidun Twarock}
\affil[1]{York Cross-disciplinary Centre for Systems Analysis, University of York, York YO10 5GE, UK}
\affil[2]{Department of Mathematics, University of York, York YO10 5DD, UK}
\affil[3]{Department of Biology, University of York, York YO10 5DD, UK}

\affil[*]{corresponding authors: ffc505@york.ac.uk (FF) and rt507@york.ac.uk (RT)}

\keywords{Key words: protein nanoparticle, interaction network, surface tiling, symmetry, AaLS pentamer cage}

\begin{abstract}
Protein nanoparticles play pivotal roles in many areas of bionanotechnology, including drug delivery, vaccination and diagnostics.~These technologies require control over the distinct particle morphologies that protein nanocontainers can adopt during self-assembly from their constituent protein components.~The geometric construction principle of virus-derived protein cages is by now fairly well understood by analogy to viral protein shells in terms of {Caspar and Klug's quasi-equivalence principle}. However, many artificial, or genetically modified, protein containers {violate this principle, because identical protein subunits do not interact in quasi-equivalent ways, leading to gaps in the particle surface}.~{We introduce a method that exploits information on the local interactions between the assembly units, called capsomers, to infer the geometric construction principle of these nanoparticle architectures.~The predictive power of this approach is demonstrated here for a prominent system in nanotechnology, the AaLS pentamer}. Our method not only rationalises hitherto discovered cage structures, but also predicts geometrically viable options that have not yet been observed.~The classification of nanoparticle architecture based on the geometric properties of the interaction network closes a gap in our current understanding of protein container structure and can be widely applied in protein nanotechnology, paving the way to programmable control over particle polymorphism. 
\end{abstract}

\begin{document}

\maketitle
\section{Introduction}

Protein containers are ubiquitous in nature. Prominent examples are the viral protein shells, called capsids, that provide protection and transport for viral genomes between rounds of infection. Protein cages also serve vital functions in bacteria as microcompartments, or in prokaryotic cells, where encapsulins, \textcolor{mycolor}{ferritin} and lumazine synthase cages facilitate catalysis \cite{Bacher2000}, intracellular trafficking \cite{Brodsky2012}, and transport \cite{Seckback1982,Szyszka2022}. Nanoparticles, either derived from these naturally occurring protein containers or \textit{de novo} designed, play pivotal roles in a host of applications, including vaccine development \cite{Balke2020,Ueda2020}, cargo storage, drug delivery, gene therapy and diagnostics \cite{de2009,Steinmetz2020}. 


Viruses have evolved mechanisms to assemble specific geometric designs with high fidelity and efficiency. The vast majority of virus architectures exhibit icosahedral symmetry as a consequence of the principle of genetic economy \cite{Crick1956}, as this symmetry type allows container volume to be maximised without increasing the coding cost for its components. The additional volume in the confines of the capsid provides room to package genes supporting other functions, thus allowing viruses to gain more complexity with time. 
\textcolor{mycolor}{Insights into the geometric and mechanical properties of viral capsids afford a better understanding of viral life cycles. For example, buckling transitions from an initial spherical procapsid to the final icosahedral faceted shell have been shown to enhance a capsid's tolerance of internal pressures \cite{Aznar2012}, and models of thermal dissociation have elucidated the processes of viral assembly and disassembly \cite{Chen2017}.} The structures of artificial nanoparticles, on the other hand, \textcolor{mycolor}{are not as well understood to date and } exhibit a much wider spectrum of morphologies \cite{Biela2022,King2012,Lai2016,Stupka2022,Majsterkiewicz2022,Sharma2022,Malay2019,Ueda2020}. \textcolor{mycolor}{This is because they frequently violate the quasi-equivalence principle. \textcolor{REFcolor}{A capsid is considered quasi-equivalent if it is held together by the same type of bonds throughout, allowing for deformations in slightly different ways in the different, non-symmetry related environments \cite{Caspar1962}.} 
For example, violation of the principle occurs if a subset of the constituent protein subunits of the assembly unit of the capsid (capsomer) does not interact with neighbouring capsomers, resulting in larger gaps in the particle surface.} As such gaps are biologically important, amongst others for diffusion-limited encapsulation of complementarily charged guest molecules \cite{Sasaki2017}, a better understanding of the geometric construction principle of such artificial protein cages is required. This is also an important step towards control over nanoparticle size and stoichiometry, enabling their manufacturing to be optimised, and their biophysical properties to be tuned for specific applications \cite{Biela2022}. 

The seminal Caspar-Klug (CK) theory was the first to propose quasi-equivalence as a geometric design principle for the structural organisation of icosahedral viruses \cite{Caspar1962}. CK theory \textcolor{mycolor}{indicates capsid protein (CP) positions relative to surface triangulations, ascribing CPs to the polyhedral angles (corners) of the triangular tiles (Fig. \ref{fig1}A).
\begin{figure}[H]
    \centering
    \includegraphics[width=0.53\linewidth]{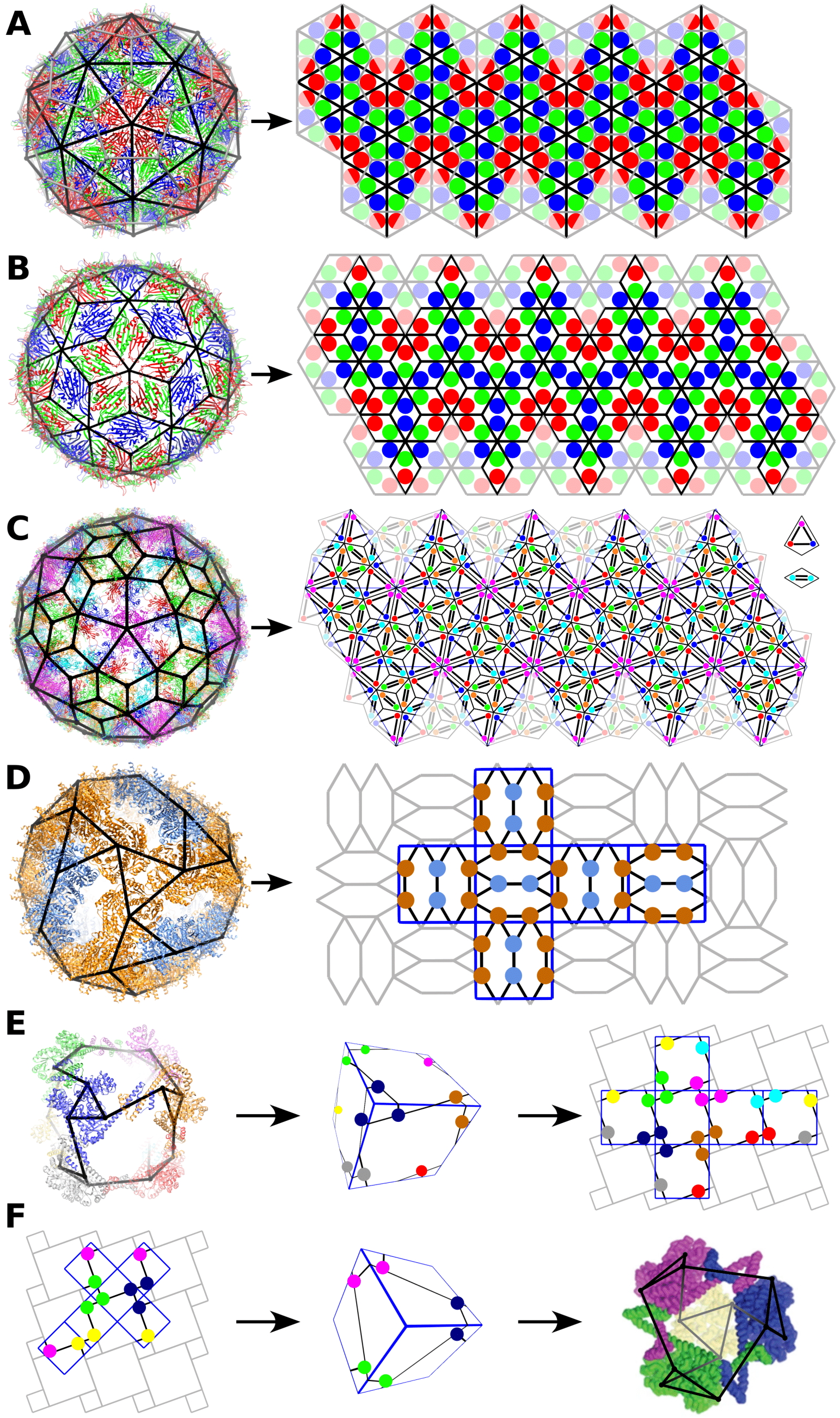}
    \caption{\textcolor{mycolor}{Tiling models of virus architecture. (A) A triangular tiling according to Caspar-Klug theory is a surface lattice model for the Pariacoto virus capsid (PDB: 1F8V); its dual, a hexagonal lattice (grey) also correctly models the relative positions of the capsid proteins. (B) A different surface lattice model, a rhomb tiling, is required to capture the relative CP positions in bacteriophage MS2 (PDB: 2MS2); rhombs are one-to-one with the protein dimers from which the capsid assembles. (C) The surface lattice of human papillomavirus (PDB: 3J6R) is made of two tiles, a kite and a rhomb, that represent trimer and dimer interactions in the capsid surface. Note that each CP interacts with CPs in other pentamers. (D) The \textcolor{REFcolor}{pentamers in} AaLS-neg, a protein cage made from 36 pentamers (PDB: 5MQ3), cannot be mapped onto \textcolor{REFcolor}{pentagons in} a surface lattice \textcolor{REFcolor}{in which every tile has an interpretation in terms of protein positions. As} a subset of CPs do not interact with proteins in other \textcolor{REFcolor}{pentamers, this cage} violat\textcolor{REFcolor}{es} the quasi-equivalence principle, result\textcolor{REFcolor}{ing} in gaps. However, the interaction network between capsomers (not protein subunits) can be represented as a tiling; its geometric information is exploited here to construct and classify alternative capsid architectures. (E) The interaction network of the protein cage reported in Ref. \cite{Lai2014} (PDB: 4QCC) can be mapped onto a cube; coloured vertices indicate the centres of mass of the CPs (middle). This cubic surface can be embedded into a gyrated square tiling (right). (F) Other embeddings of the cubic surface into the gyrated square tiling (left) predict the morphologies of other cages that can assemble from the same protein units, such as the smaller cage reported in  \cite{Lai2014,Lai2016} (right).}}
    \label{fig1}
\end{figure}
\noindent
The dual tilings, polyhedra with hexagonal and pentagonal faces akin to Buckminster Fuller's domes \cite{Marks1960},  predict the same capsid layout, again assuming CPs to be located in the polyhedral angles of the faces (Fig. \ref{fig1}A, grey). Therefore viral capsids are interchangeably modelled via hexagonal surface lattices and triangulations in CK theory. These models predict protein positions correctly for capsids formed from pentagonal, hexagonal or triangular capsomers, i.e. from pentamers, hexamers and trimers. However, they do not accurately reflect the layout of capsids assembled from dimers (SI Fig. S1). Viral Tiling theory (VTT) recognises that this is because the tiles in the surface lattice must be in a one-to-one correspondence with biological units. It therefore represents bacteriophage MS2, which assembles from protein dimers, by a rhomb tiling (Fig. 1B), thus capturing the correct relative CP orientations. More general types of tilings are required for other capsomer types if viral capsids are formed from more than one type of protein unit \cite{Twarock2019}.}

\textcolor{mycolor}{A further complication arises for human papillomavirus (HPV), a capsid formed from 72 identical pentamers. 
Due to the crystallographic restriction \cite{Senechal1996} there is no all-pentamer surface lattice with more than 12 identical pentagonal tiles, implying that pentamers cannot be modelled \textcolor{REFcolor}{in a biologically meaningful way} by pentagonal tiles. Again recognising the need for tiles to have an interpretation in terms of biological units, VTT introduces two types of tiles for HPV, a rhomb and a kite, that are in a one-to-one correspondence with the two types of interactions \textcolor{REFcolor}{mediated by the C-terminal arms} stabilising the capsid: kites corresponding to three proteins forming a trimer interaction, and rhombs representing two proteins forming a dimer interaction \cite{Twarock2004} (Fig. \ref{fig1}C). This tiling is reminiscent of the Penrose tiling \cite{Penrose1979}, an aperiodic tiling given in terms of kites and rhombs, and 3D Penrose tilings can also be used to approximate virus structure \cite{Salthouse}. There are other approaches modelling virus architecture, using a local rules approach \cite{Schwartz2000} or 
a dodecahedral nets approach that associates the centres of mass of protein molecules in icosahedrally symmetric viral cages with the nodes of a chiral pentagonal tiling  \cite{Konevtsova2017}. However, the lack of a direct correspondence between tiles and biological units limits the predictive power of these approaches. They describe the layouts of different viruses built according to the same mathematical principle, but do not allow for the classification of particle morphologies that can be formed from the same types of building blocks. By contrast, this is possible in VTT as tiles have an interpretation in terms of assembly units or specific types of protein-protein interactions.}

Protein nanoparticles exhibit a wider spectrum of morphologies than viruses because they also include structures violating the quasi-equivalence principle. This occurs if some of the protein subunits in a capsomer do not interact with other capsomers, as is the case for the 36-pentamer particles formed from \textit{Aquifex aeolicus} lumazine synthase (AaLS) \cite{Sasaki2017}(Fig. \ref{fig1}D). In this case, neither capsomers nor interactions between individual CPs can be represented by tiles in a meaningful way. Of course, the particle surface can always be represented as a tessellation in terms of multiple copies of the fundamental domain (also called asymmetric unit) of the underlying symmetry group, in this case consisting of three Voronoi cells (SI Fig. S2). However, in such mathematical representations, there is no clear biological interpretation of the lattice unit in terms of individual capsomers or interactions between protein subunits. Therefore, such an approach does not allow prediction of other possible cages made from AaLS pentamers.

\textcolor{mycolor}{In order to achieve this, we construct the interaction network between capsomers (rather than between their constituent protein subunits) \cite{Brunk2021}. For this, the centres of mass (CoMs) of the capsomers are computed based on the coordinates in the PDB file. These then form nodes in a network in which connecting edges indicate interactions between capsomers (Fig. \ref{fig1}D). This coarse-grained topological descriptor of capsid architecture ignores the geometry of individual capsomers, and interactions formed by individual protein subunits, and is therefore distinct from the surface lattice models in VTT. However, the geometric structure of this interaction network can be embedded into a tiling. For this, the symmetry axes of the particle are aligned with those of a reference cube (see also SI Fig. S3), and then the cubic surface is embedded into a planar tiling that continues the interaction network periodically in the plane. This tiling can then be used to construct models for other particle types via different embeddings of the cubic surface, akin to the embedding of icosahedral surfaces into hexagonal lattices in Caspar-Klug theory.}

\textcolor{mycolor}{The cubic protein container designed by Lai et al. \cite{Lai2014} (Fig. \ref{fig1}E, left) provides a simple example of this interaction network approach. Representing each of its 24 proteins as a node ( Fig. \ref{fig1}E, left) and drawing connections between interacting proteins, results in the interaction network (shown in black). By aligning the 4-fold and 3-fold symmetry axes of the particle with those of a reference cube (Fig. \ref{fig1}E, middle), the interaction network can be mapped onto the cubic surface. The latter is then embedded into a planar tiling by ``unfolding'' the cubic surface in the plane (Fig. \ref{fig1}E, right). Any other particles assembled from the same protein units should exhibit similar local interactions. From a mathematical point of view, this means that their interaction networks can be constructed by working backwards from the planar tiling. In particular, any other planar embedding of the cubic surface, obtained via rescaling and reorienting the surface in the plane such that the symmetry axes of the cube again coincide with those of the tiling (see also SI Fig. S3), then present an alternative particle layout. To reconstruct the biological model, vertices have to be replaced by biological units, oriented such that interacting units meet along the edges of the interaction network (Fig. \ref{fig1}F). This cage architecture, inferred via our method, has been observed \cite{Lai2014,Lai2016}, suggesting that our method can indeed be used to predict viable alternative protein container designs.} 

\textcolor{mycolor}{The ability to predict alternative particle morphologies that can assemble from the same protein unit(s) is important in nanotechnology. It can be used in the context of kinetic models to compute relative ratios of different particle morphologies for different experimental conditions, thus opening up the opportunity to tune experiments to favour production of desired particle types \cite{Biela2022}. It also informs the reconstruction of less frequent particle types from cryo-EM data in the case of polymorphic assembly, and guides the selection of particle morphologies with desired biophysical properties for specific applications \cite{Brunk2021}.} In the following, we demonstrate the predictive power of our approach for a more complex system -- cages formed from AaLS pentamers -- in which nodes in the interaction network represent assembly units composed of multiple protein subunits. This analysis illustrates how the interaction network approach can be used to predict and classify particle structures for any system of interest in bionanotechnology.



\section{The building blocks of the interaction network}
\begin{figure}[H]
    \centering
    \includegraphics[width=0.7\linewidth]{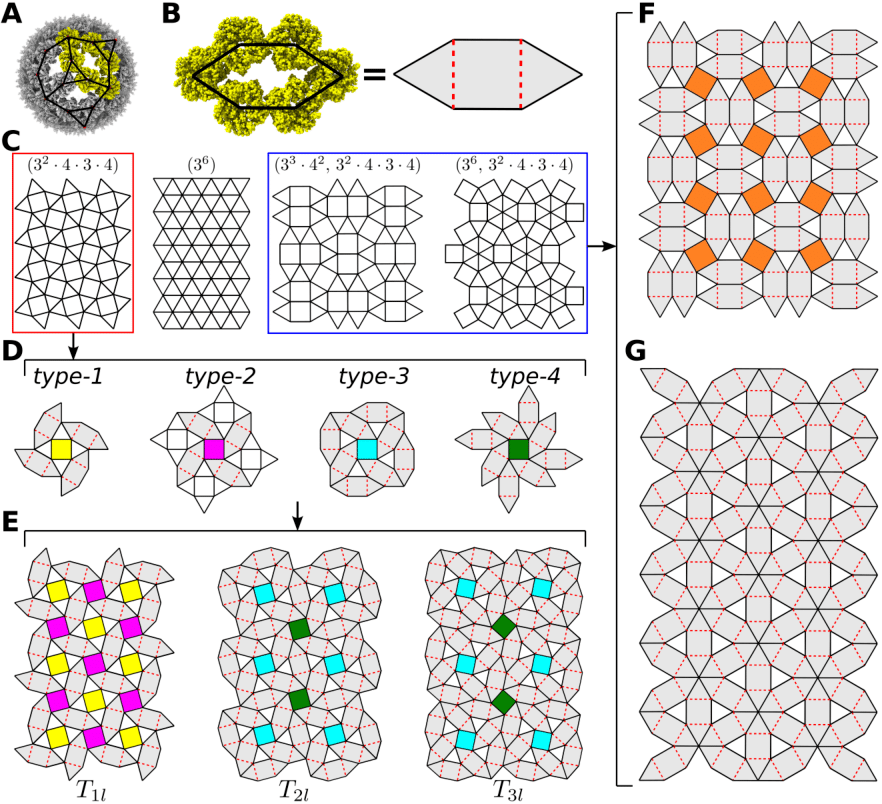}
    \caption{Classification of AaLS protein container architectures based on the interaction network approach. (A) The AaLS-13 cage, made of 72 pentamers (PDB: 5MQ7), reveals two types of interactions between pentamers: in groups of three (triangles) and groups of six (squashed hexagons). (B) Close-up view of a squashed hexagon and its schematic representation in terms of two triangles and one square; dashed red lines are used to divide the squashed hexagon into a square and two triangles. (C) The only $k$-uniform tilings (up to $k=5$) that can be partitioned into triangles and squashed hexagons. (D) There are four distinct ways in which a 4-fold symmetry axis in the snub square tiling can be surrounded by triangles and squashed hexagons. (E) There are only three types of tilings, modulo handedness, that can be constructed from these vertex environments: $T_{1l}$ from \textit{type-1} and \textit{type-2}, and $T_{2l}$ and $T_{3l}$ from \textit{type-3} and \textit{type-4}. (F) and (G) The unique way in which the ($3^3\cdot 4^2$, $3^2\cdot 4\cdot 3\cdot 4$) and ($3^6$, $3^2\cdot4\cdot3\cdot4$) tilings can be partitioned into triangles and squashed hexagons, respectively.}
    \label{fig2}
\end{figure}

Given a nanoparticle of interest, the first step consists in computing its interaction network. In AaLS-based nanoparticles, pentamers and their genetic variants self-assemble into a spectrum of different particles with tetrahedral and icosahedral symmetry \cite{Sasaki2017}. Representing the CoMs of the pentamers as nodes and drawing edges between interacting pentamers, we obtain the interaction network (Fig. \ref{fig2}A).
\noindent
The next step in the analysis is to identify the distinct geometric shapes in the interaction network. \textcolor{mycolor}{In this case, there are two types: interactions in groups of three, corresponding to triangles, and interactions in groups of 6, corresponding to non-regular hexagons. The latter, called squashed hexagons in the following because of their characteristic shapes, can be divided into two triangles and one square as indicated by dashed red lines in Fig. \ref{fig2}B}. 

\section{Planar tilings representing the interaction network}
The interaction network of AaLS pentamer cages can therefore be embedded into planar tilings made of triangles and squashed hexagons. For other nanoparticles, the nature of tiles will depend on the local interaction patterns and may therefore differ. \textcolor{mycolor}{However, as nanoparticles self-assembling from (potentially multiple different) protein units exhibit only a limited spectrum of distinct local interaction patterns between neighbouring capsomers, planar tilings associated with their interaction networks must all be $k$-uniform tilings, meaning they are tessellations of the plane with only a limited number ($k$) of distinct vertex types (cf. SI Fig. S4 and SI text). In the case of AaLS-based nanoparticles, we therefore consider $k$-uniform tilings made of triangles and squares as an intermediate step to obtaining all possible tilings in terms of triangles and squashed hexagons by deleting edges in the triangle-square tilings.}

\section{Classification of tilings associated with the interaction network}
\textcolor{mycolor}{$k$-uniform tilings have been classified. There are in total 575 tilings with polygonal faces up to $k=5$ (see SI text),} 140 of which are made entirely of triangles and squares \cite{Chavey1989,Grunbaum1987,Galebach}. The combinatorial task is to identify all possible tilings that are given exclusively in terms of the building blocks of the interaction network, here triangles and squashed hexagons. \textcolor{mycolor}{Note that the tiling can potentially also include a set of squares aligning with the 3-fold symmetry axes of the cage, as these become triangles in the 3D surface (see Fig. \ref{fig1}E and F for an example).} As tilings in which four squares tessellate a bigger square cannot be subdivided into triangles and squashed hexagons, these tilings are therefore excluded from further analysis, reducing the number of candidates to 58. In order to construct a particle with tetrahedral or octahedral symmetry from such tilings, the symmetry axes of the protein cage must be identified with symmetries in the tiling, as illustrated in Figs. \ref{fig1}E and F. This requires the tiling to have 3-fold and/or 4-fold symmetries. 46 of the 58 tilings have only 2-fold symmetry and therefore cannot be used to construct the surface lattices of particles with tetrahedral or octahedral symmetry. We checked each of the 12 remaining tilings individually, excluding five tilings (SI Fig. S5A) from further consideration (cf. SI Fig. S6A and SI text). 
Three of the remaining seven tilings (SI Fig. S5B) contain local 6-fold symmetry axes and therefore must also be excluded as they would require vertices representing pentamers to be positioned on a 6-fold symmetry axis (cf. SI Fig. S6B and SI text).
\textcolor{mycolor}{In summary, only four tilings fulfil all necessary criteria to allow for the embedding of an AaLS interaction network. These are the triangular ($3^6$) and the snub square ($3^2\cdot4\cdot3\cdot4$) tiling, which are both uniform, and the two 2-uniform tilings ($3^3\cdot 4^2$, $3^2\cdot 4\cdot 3\cdot 4$) and ($3^6$, $3^2\cdot 4\cdot 3\cdot 4$) (Fig. \ref{fig2}C).}

\textcolor{mycolor}{For each of these tilings, we next identify all possible ways in which they can be reorganised into triangles and squashed hexagons. First, we focus on the snub square tiling which has 4-fold symmetry axes at the centres of its squares (Fig. \ref{fig2}C, left)}. There are four inequivalent options of organising triangles and squashed hexagons around these axes (Fig. \ref{fig2}D). One option is to accommodate four triangles around the square (yellow) and continue  with squashed hexagons (\textit{type-1}). The other option is to organise four squashed hexagons around a square (magenta) and then continue in one of the following ways: either locate triangles between the squashed hexagons (\textit{type-2}) or accommodate  the squashed hexagons in a way to either connect (\textit{type-3}) or place them between (\textit{type-4}) the previously added hexagons.

Starting with a \textit{type-1} square, the only option is to continue the tiling with \textit{type-2} squares (SI Fig. S7A), and vice versa, resulting in the tiling called $T_{1l}$ (Fig. \ref{fig2}E) as the unique solution. Starting from a \textit{type-3} square, there are two squares where choices have to be made (SI Fig. S8A). In each case, a \textit{type-4} square is required next, and then the tessellation must be continued with alternating \textit{type-3}/\textit{type-4} squares (SI Fig. S8B and C), leading to the tilings $T_{2l}$ and $T_{3l}$ (Fig. \ref{fig2}E). The same construction can be applied to the right-handed versions of the squares, i.e. the opposite handed versions obtained  using the mirror images of the configurations in Fig. \ref{fig2}D. This results in analogous right-handed tilings, denoted as $T_{1r}$ (SI Fig. S7B), $T_{2r}$ and $T_{3r}$ (SI Fig. S8D). Thus, modulo handedness, the snub square tiling can be divided into squashed hexagons and triangles in precisely three inequivalent ways, corresponding to the tilings $T_{jl}$ ($j=1,2,3$) shown in Fig. \ref{fig2}E. 

For all other tilings, the combinatorics are much simpler. The ($3^3\cdot 4^2$, $3^2\cdot 4\cdot 3\cdot 4$) tiling can be divided into triangles and squashed hexagons in only one way (Fig. \ref{fig2}F). \textcolor{mycolor}{In the ($3^6$, $3^2\cdot 4\cdot 3\cdot 4$) tiling, placement of the 6-fold vertices must be such that the particle generated from the tiling does not contain any 6-fold symmetric vertices as this would be incompatible with vertices representing pentamers. Thus, this tiling can only be subdivided into triangles and squashed hexagons in a unique way (Fig. \ref{fig2}G).}


\section{Construction of protein cage architectures}
\begin{figure}[H]
    \centering
    \includegraphics[width=0.7\linewidth]{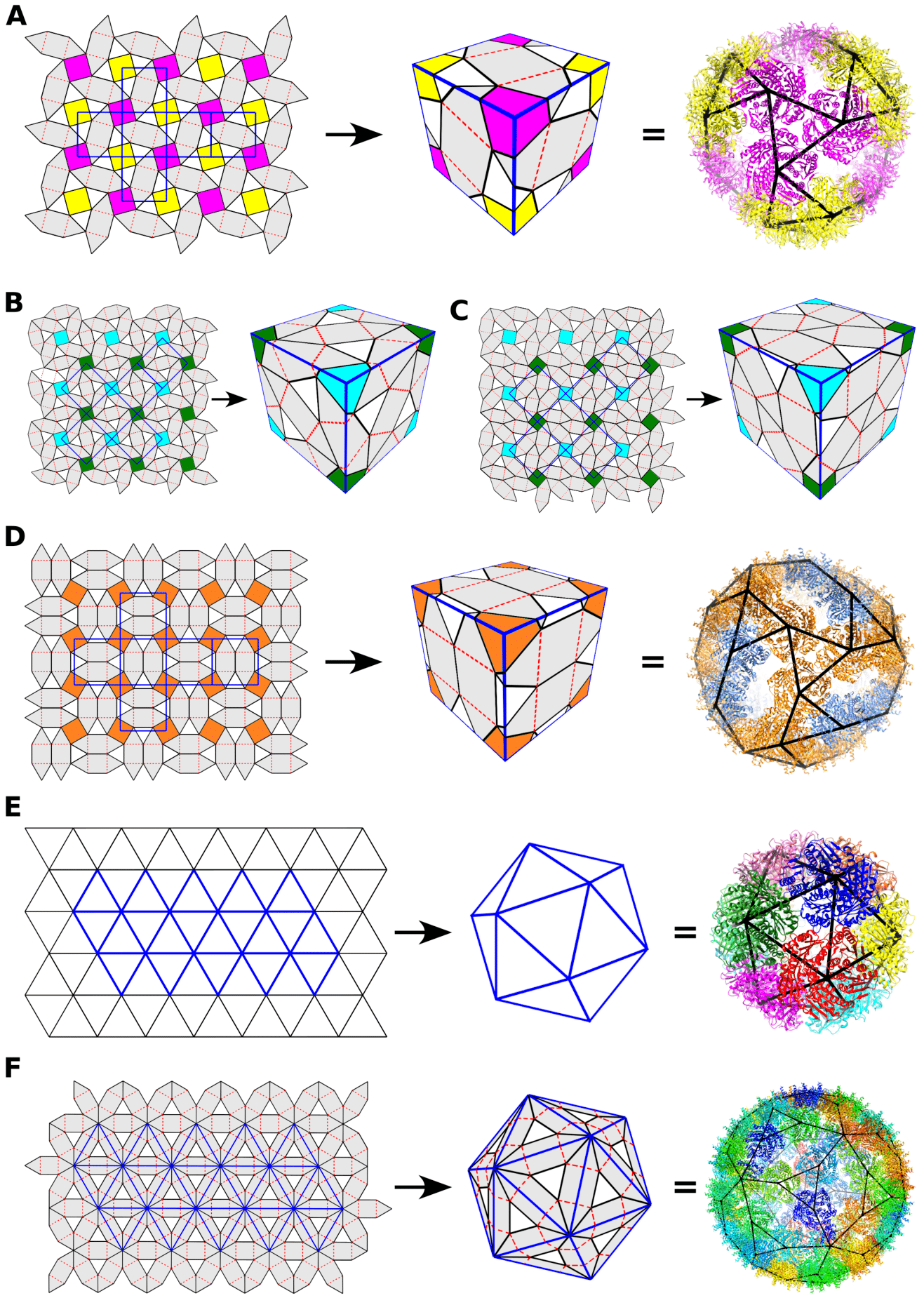}
    \caption{AaLS cage architectures derived from interaction networks. (A) The $T_{1l}$ tiling provides the layout of a particle made from 24 pentamers and corresponds to the surface structure of a known AaLS protein cage (PDB: 7A4F). (B) and (C) Particle architectures derived from the $T_{2l}$ and $T_{3l}$ tilings correspond to cages with  tetrahedral symmetry made from 48 and 60 pentamers, respectively. These cages have not been reported to date. (D) The ($3^3\cdot 4^2$, $3^2\cdot 4\cdot 3\cdot 4$) tiling predicts a particle made from 36 pentamers with tetrahedral symmetry. Its surface architecture corresponds to a known AaLS protein cage (PDB: 5MQ3). (E) The triangular tiling corresponds to a polyhedron with 12 vertices, which embodies the architecture of the WT AaLS cage (PDB: 5MPP). (F) The ($3^6$, $3^2\cdot 4\cdot 3\cdot 4$) tiling corresponds to an  icosahedral particle made from 72 pentamers, and corresponds to the AaLS-13 protein cage (PDB: 5MQ7).}
    \label{fig3}
\end{figure}

Given the exhaustive list of tilings embodying the characteristics of the interaction network derive above, models for particles with tetrahedral or octahedral symmetry can then be constructed via different embeddings of a cubic surface (see SI Fig. S3). For this, a planar representation of the cubic surface must be embedded into the tiling such that its corners align  with the centres of appropriately spaced squares in the tiling. For example, allocating the vertices of the cubic surface to the centres of adjacent coloured squares in the $T_{1l}$ tiling (Fig. \ref{fig3}A, left), and mapping these onto the vertices of a cube (Fig. \ref{fig3}A, middle), generates a model for an AaLS cage made of 24 pentamers (Fig. \ref{fig3}A, right). This particle has been observed in the self-assembly of AaLS pentamers \cite{Tetter2021}. 

It is not possible to embed the surface of a cube into this tiling in any other way without mapping squares in the tiling onto the faces of the cube (cf. SI Fig. S9A). As vertices represent pentamers, this would generate a ring-like interaction between four AaLS pentamers in the particle surface, which is a local  interaction type that has not been observed in any AaLS cage to date. We therefore exclude it from our classification. There are thus no other biologically viable AaLS cage models that can be constructed from this tiling. 
Similarly, both $T_{2l}$ and $T_{3l}$ result in only one tetrahedral model each.  The former corresponds to a protein cage made from 48 pentamers (Fig. \ref{fig3}B), and the latter to a 60 pentamer-cage (Fig. \ref{fig3}C). These particles have not yet been observed but are consistent with the experimentally observed local interaction rules and therefore provide viable geometric models for AaLS cages. It is possible that these cages have previously been overlooked because they occur less frequently during polymorphic assembly than other variants. Note that all known AaLS cages only exhibit the left-handed version of the tilings. We therefore only consider the left-handed versions in our analysis, assuming that the interactions in all cages should have similar characteristics. 

The 2-uniform tiling ($3^3\cdot 4^2$, $3^2\cdot 4\cdot 3\cdot 4$) has only one type of 4-fold symmetry axis, shown in orange (Fig. \ref{fig3}D). In analogy to the snub square tiling, only the smallest embedding of the cubic surface is possible, and is obtained by associating neighbouring squares with the vertices of a cubic face (Fig. \ref{fig3}D). This particle is made from 36 pentamers that are organised with tetrahedral symmetry, and corresponds to one of the AaLS pentamer cages reported previously \cite{Sasaki2017}. Note that, as before, any larger models, obtained via an embedding of a rescaled cubic surface (SI Fig. S9) would necessarily contain a square, i.e. a group of four pentamers, and we again reason that this would not be a biologically viable option.

The remaining two tilings include 6-fold symmetry axes. It is therefore possible to construct particle architectures with icosahedral symmetry from them following the Caspar-Klug construction \cite{Caspar1962}. The smallest particle with icosahedral symmetry that can be derived from the triangular tiling is the icosahedron, a polyhedron with 12 vertices. This model corresponds to the wild-type (WT) AaLS cage (Fig. \ref{fig3}E). In the Caspar-Klug construction, higher order triangulations are possible in which the faces of the icosahedron are subdivided into triangular facets. However, in their surface lattice interpretation, proteins are allocated in the $60^{\circ}$ angles of the triangular facets, thus generating models with 12 pentamers and otherwise hexamers. Here, on the other hand, the triangulation is representing the interaction network between pentamers, whose positions are indicated by the vertices. Therefore, larger particles with icosahedral symmetry are not feasible as they would map pentamers onto 6-coordinated vertices in the triangulation. 
Similarly, tetrahedral and octahedral particles can be constructed via triangular surface lattices \cite{Twarock2019}, but are not viable in the framework of AaLS interaction networks as they would locate pentamers on 3- or 4-fold symmetry axes. Similar arguments show that there exists only one planar embedding of an icosahedral surface into the ($3^6$, $3^2\cdot 4\cdot 3\cdot 4$) tiling leading to a viable AaLS cage (Fig. \ref{fig3}F). The resulting cage morphology, a particle made of 72 pentamers, corresponds to the AaLS-13 cage \cite{Sasaki2017}.

There are thus precisely six cage structures with 3D symmetry that can self-assemble according to the known local interaction pattern of the AaLS pentamers (Fig. \ref{fig3} and Table \ref{table1}), \textcolor{mycolor}{ranging in size from the WT AaLS cage (12 pentamers) to the AaLS-13 cage (72 pentamers)}. \textcolor{mycolor}{Our classification implies that there are no AaLS particles with octahedral symmetry.} It also predicts intermediate structures formed from 48 and 60 pentamers that could potentially be observed but have not been reported to date. \textcolor{mycolor}{Note that the distinct symmetry types observed here for AaLS cage assembly can also occur in the self-assembly of other protein cages \cite{Panahandeh2018,Paquay2016}.} Our analysis demonstrates how tilings encoding the interaction network of a protein cage can be used to systematically enumerate all possible alternative protein cage structures with 3D symmetry that can also assemble from the same protein units. Whilst we have demonstrated our approach for the AaLS system, it can readily be applied to any protein nanocontainer architecture of interest following the methodology introduced here. This analysis is thus a primer for the modelling of protein nanocontainers in bionanotechnology. 

\begin{table}
\caption{Classification of AaLS protein cages with cubic symmetry \cite{Crick1956}}
\label{table1}
\centering
\begin{tabular}{lclll}
\hline
Tiling                                                   & No. of pentamers & Symmetry    & Experimental observation & Figure \\\hline\hline
Triangular ($3^6$)                                       & 12   & Icosahedral & Observed (PDB: 5MPP \cite{Sasaki2017}) & Fig. \ref{fig3}E               \\\hline
Snub square tiling ($3^2\cdot 4\cdot 3\cdot 4$) $T_{1l}$ & 24   & Tetrahedral & Observed (PDB: 7A4F \cite{Tetter2021}) & Fig. \ref{fig3}A                \\\hline
($3^3\cdot 4^2, 3^2\cdot 4\cdot 3\cdot 4$)               & 36   & Tetrahedral & Observed (PDB: 5MQ3 \cite{Sasaki2017}) & Fig. \ref{fig3}D                \\\hline
Snub square tiling ($3^2\cdot 4\cdot 3\cdot 4$) $T_{2l}$ & 48   & Tetrahedral & Not observed & Fig. \ref{fig3}B            \\\hline
Snub square tiling ($3^2\cdot 4\cdot 3\cdot 4$) $T_{3l}$ & 60   & Tetrahedral & Not observed & Fig. \ref{fig3}C            \\\hline
($3^6, 3^2\cdot 4\cdot 3\cdot 4$)                        & 72   & Icosahedral & Observed (PDB: 5MQ7 \cite{Sasaki2017}) & Fig. \ref{fig3}F \\\hline               
\end{tabular}
\end{table}

\section{Discussion}


Protein containers, either adapted from naturally occurring protein cages or {\it de novo} designed, are pillars of bionanotechnology. Many groups worldwide are developing novel types of nanoparticles for a host of applications, for example using the Rosetta Software \cite{Ueda2020,Divine2021,Hsia2021,Cannon2020}. The simultaneous assembly of a wide spectrum of particle morphologies -- a phenomenon known as particle polymorphism -- poses a challenge for nanocontainer production. \textcolor{mycolor}{Such polymorphism, which has also been observed in the assembly of capsid proteins in the presence and absence of viral RNA genomes \cite{Bond2020,Zandi2020}, is often triggered by genetic modifications of the capsid protein subunit. This includes insertion of amino acid sequences (SpyTags) into the outward facing portion of the protein subunits, a method that is standardly used to functionalise the particle surface (using SpyCatchers) with antigens for vaccine production. Such modifications are known to result in the assembly of multiple different particle morphologies that contain the WT morphology as only one of many distinct options \cite{Biela2022}. Genetic modifications to alter the chemical properties of the protein units, such as their charges or sensitivity to pH, have similar effects, both in nanocontainers derived from bacterial enzymes   \cite{Sasaki2017,Aumiller2018,Azuma2018,Koziej2022} and in virus-like particles  \cite{Bond2020,Panahandeh2022}. Understanding the determinants of this particle polymorphism is an important step in controlling the assembly outcome.} 

\textcolor{mycolor}{We introduce here a theoretical framework to characterise the spectrum of nanoparticle morphologies that can assemble from a given set of protein units. This interaction network approach can be used even for container architectures violating the principle of quasi-equivalence, that is central to the seminal CK theory \cite{Caspar1962}, VTT \cite{Twarock2004,Keef2005,Keef2006,Keef2008,Elsawy2008,Twarock2019,Twarock2005m} and related models explaining the surface architectures of {\it de novo} designed nanoparticles used as malaria vaccines \cite{Indelicato2017}. These approaches fail, because the existence of protein subunits not interacting with neighbouring capsomers makes it difficult to define a biologically meaningful mathematical unit for the tiling models.}
The method introduced here closes this gap in our understanding of protein nanocontainer architecture. It uses knowledge of the local interactions between the self-assembling capsomers to systematically enumerate all viable container designs with 3D symmetry that can be formed from them. The predictive power of this approach is demonstrated for particles formed from AaLS pentamers  \cite{Sasaki2017,Tetter2021}, for which our method not only characterises all experimentally observed variants of different sizes and 3D symmetries, but also predicts structures that have not been observed yet. \textcolor{mycolor}{As tiles in CK theory and VTT represent capsomers, the dual tilings -- obtained by replacing tiles by vertices and connecting vertices corresponding to adjacent tiles -- can also be viewed as interaction networks (SI Fig. S10). Thus, the interaction network approach provides a unifying framework for the modelling of virus and nanoparticle architecture alike. It is applicable in bionanotechnology for the classification of nanocontainer architecture and can be built as local constraints into programmes like Rosetta to support nanoparticle design. The geometric models also open up novel avenues for the study of the biophysical properties of protein cages, such as their  propensity for fragmentation and cargo release \cite{Brunk2021}, or their kinetics of self-assembly along assembly pathways leading to distinct particle types \cite{Biela2022}. Such models predict the relative ratios of different particle types depending on experimental conditions, revealing how assembly can be biased towards specific outcomes. This, in turn, provides a means to increase the yield of desired particles, supporting the rational design of protein containers for diverse applications in bionanotechnology.}

\section*{Data availability}

All study data are included in the article and/or Supplementary Information.

\section*{Acknowledgements}

RT thanks the Wellcome Trust for financial support through the Joint Investigator Award (110145 \& 110146), the EPSRC for an Established Career Fellowship (EP/R023204/1) which also provided funding for FF and the Royal Society for a Royal Society Wolfson Fellowship (RSWF/R1/180009). 

\section*{Author contributions statement}

The authors made equal contributions in the article. All authors read and approved the final manuscript.

\section*{Competing interests}

The authors declare no conflict of interest.


\newpage


{\centering
    \Huge\bf Supplementary material \\\vspace{2cm}
}

\renewcommand{\thefigure}{S\arabic{figure}}
\renewcommand{\thesection}{S\arabic{section}}

\setcounter{figure}{0}
\setcounter{section}{0}

\section{Tiling models of protein container architecture - definitions and concepts}

The first tiling models of viral protein containers were introduced by Caspar and Klug. They used hexagonal lattices and triangulations to pinpoint the positions of capsid proteins in quasi-equivalent positions, i.e. occupy similar local environments in the capsid shell. These lattices are special cases of uniform tilings: A \textbf{uniform tiling} is a tessellation of the plane by convex regular polygons that is vertex-transitive, i.e., all of its vertices are equivalent under the symmetries of the tiling. 

There are 11 uniform tilings, also called Archimedean lattices. As all vertices of a uniform tiling are surrounded by identical arrangements of polygons, this feature is used to label different options. For example, the Kagome lattice (Fig. \ref{SI4}A) is a uniform tiling in which each vertex is surrounded in clockwise or anti-clockwise order by a triangle, hexagon, triangle and hexagon. It is therefore denoted as ($3\cdot6\cdot3\cdot6$), where numbers refer to the numbers of corners in each polygon in successive order. 

In the context of the interaction network approach, we also allow for capsid architectures that violate the quasi-equivalence principle. In this case, there can be more than one type of vertex environment, and we therefore need $k$-uniform tilings: A \textbf{$k$-uniform tiling} of the plane is a tessellation by convex regular polygons, connected edge-to-edge, with $k$ distinct types of vertex environments. In analogy to uniform tilings, they are labelled by $k$ indices characterising their distinct vertex environments. For example, the vertices of the 2-uniform tiling in Figure \ref{SI4}B are either surrounded by three triangles and two squares (type ($3^3\cdot 4^2$), black disk), or two triangles, a square, a triangle and a square (type ($3^2\cdot 4\cdot 3\cdot 4$), grey disk), and this tiling is therefore labelled as ($3^3\cdot 4^2$, $3^2\cdot 4\cdot 3\cdot 4$). $k$-uniform tilings have been enumerated, and there are 20 2-uniform tilings, 61 3-uniform tilings, 151 4-uniform tilings and 332 5-uniform tilings, thus 575 in total up to $k=5$.

\color{black}

\section{Arguments for the exclusion of tilings from the classification}

The classification of AaLS surface lattices relies on the identification of $k$-uniform tilings that can be partitioned into triangles and squashed hexagons. Starting with all 140 tilings given in terms of triangles and squares, we derive tilings in terms of triangles and squashed hexagons (the characteristic local interactions of AaLS cages) by deleting edges. Here we elaborate on two arguments that are used to exclude specific types of $k$-uniform tilings (cf. main text p.4):

\textbf{Exclusion of tilings that cannot be partitioned into triangles and squashed hexagons}: Figure \ref{SI5}A shows tilings that cannot be partitioned exclusively into triangles and squashed hexagons and are therefore excluded from our analysis. Figure \ref{SI6}A illustrates why this is the case based on the example of the 3-uniform tiling ($3^6$, $3^3\cdot4^2$, $3^2\cdot4\cdot3\cdot4$). The square marked by a star cannot be the centre of a 4-fold symmetry axis, and a squashed hexagon can be placed on it in only one way as shown in grey. Now the square marked by a disk cannot be the centre of a 4-fold symmetry axis, and no squashed hexagon can be placed on it, thus this tiling cannot be converted into a tessellation by triangles and squashed hexagons via deletion of edges. An similar argument excludes all five tilings in Fig.\ref{SI5}A.

\textbf{Exclusion of tilings containing local 6-fold symmetry axes}: Figure \ref{SI5}B shows tilings that can be divided into triangles and squashed hexagons, but contain local 6-fold symmetry axes. As vertices in the tilings represent pentamers, and pentamers cannot be placed on a 6-fold symmetry axis, these tilings are also excluded. Fig. \ref{SI6}B illustrates this for the 5-uniform tiling ($3^6$, ($3^2\cdot4\cdot3\cdot4$)4): Even though it is possible to identify 4-fold symmetry axes (Fig. \ref{SI5}B, yellow squares) that could align with those of a cube, even the smallest square face would necessarily contain a local 6-fold symmetry axis (red dot), which is not compatible with the interpretation of vertices as locations of pentamers. Similar arguments also apply to the other two tilings in Fig. \ref{SI5}B.

\newpage

\begin{figure}[H]
	\centering
	\includegraphics[width=0.5\linewidth]{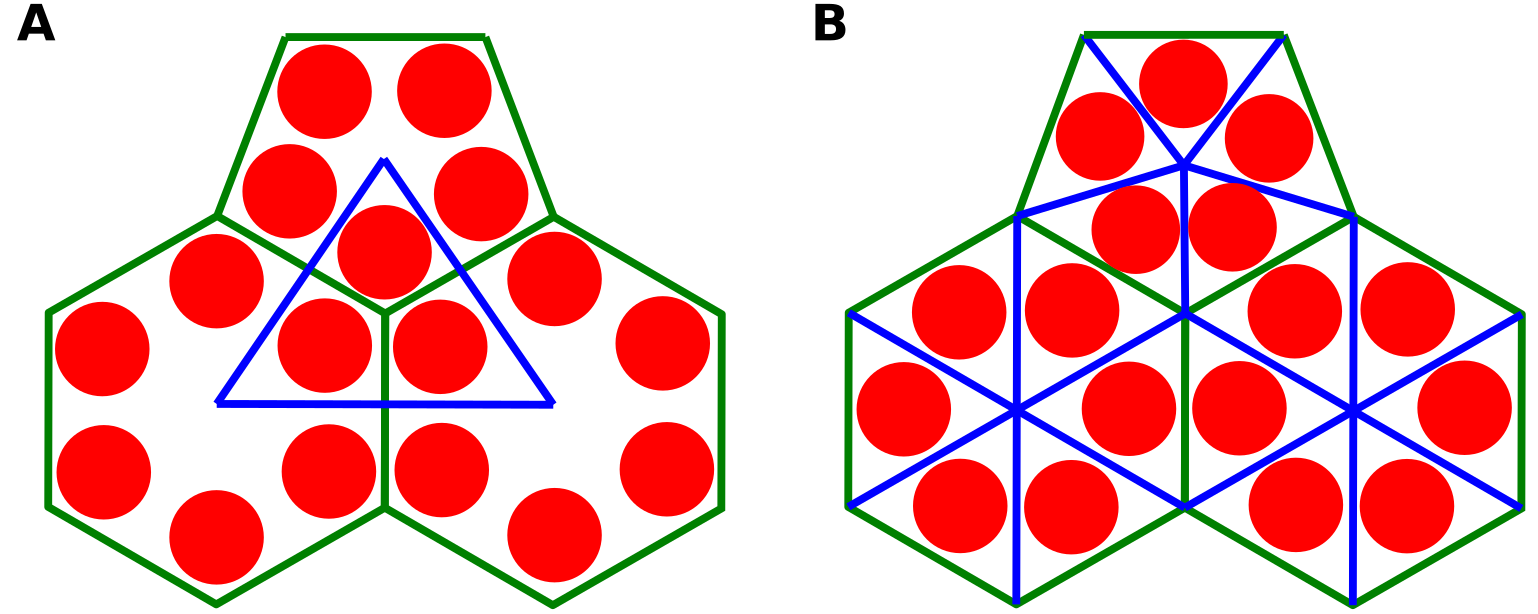}
	\caption{{Distinct tiling types predict different orientations of the capsomers (assembly units, usually composed of several protein subunits) in the capsid surface. A triangulation modelling Pariacoto virus (A), and a rhomb tiling representing bacteriophage MS2 (B), predict different relative positions of the protein subunits (red dots). Locating protein positions in the corners of the triangular or rhomb tiles (blue) following the convention in Caspar-Klug theory, the resulting capsid blueprints have different orientations with respect to the underlying pentagonal/hexagonal lattice architecture. In the triangulation pairs of protein subunits, and in the rhomb tiling individual protein subunits, in neighbouring hexamers are facing each other.}}
	\label{SI1}
\end{figure}
\newpage
\begin{figure}[H]
	\centering
	\includegraphics[width=0.9\linewidth]{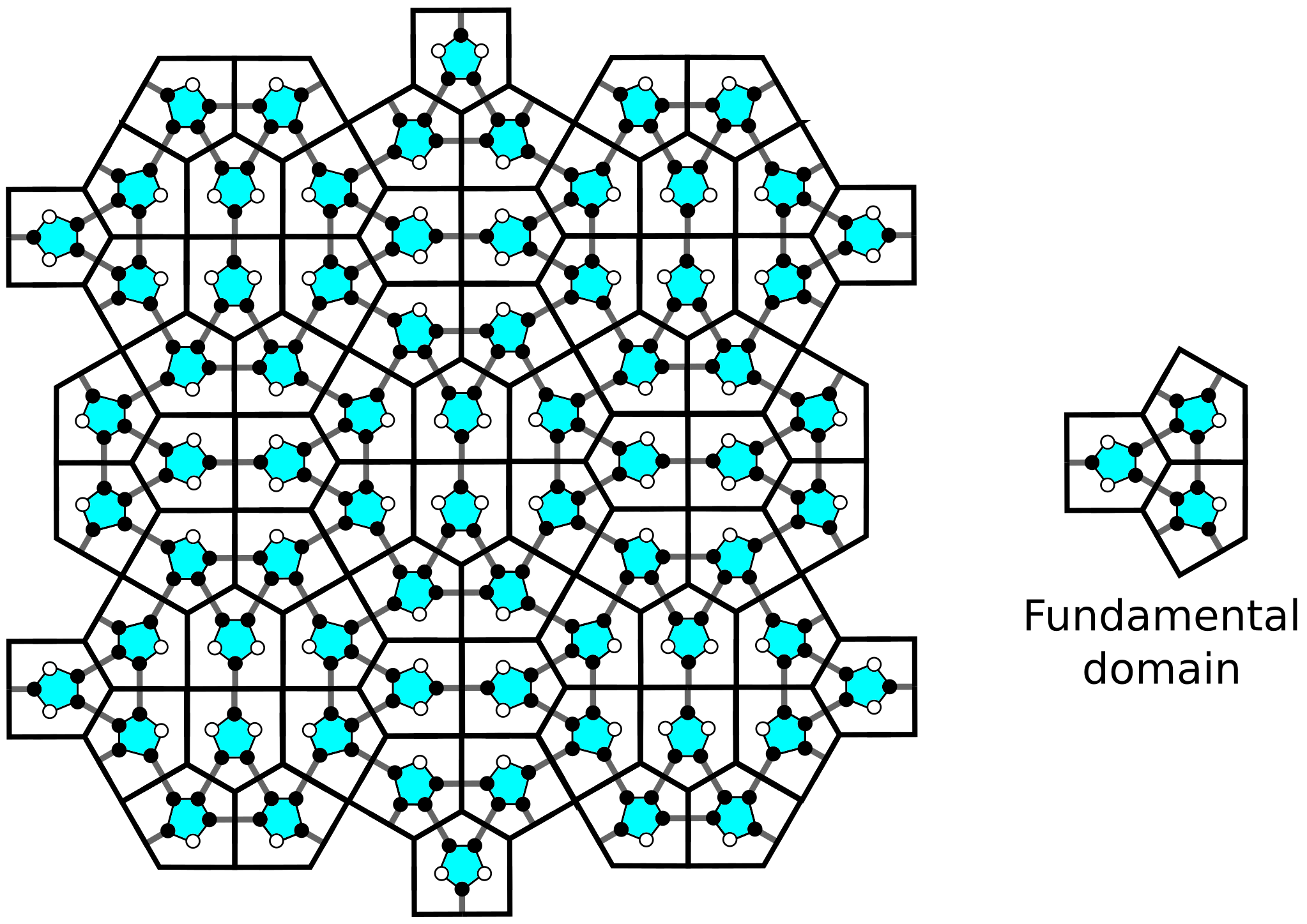}
	\caption{{Surface tessellation of the 36-pentamer AaLS cage architecture in terms of Voronoi cells. The fundamental domain (or asymmetric unit) of the tiling consists of three Voronoi cells (right). Pentamers are coloured in cyan, and black and white circles indicate monomers that bind, and respectively do not bind, to a protein subunits of an adjacent pentamer.}}
	\label{SI2}
\end{figure}
\newpage
\begin{figure}[H]
	\centering
	\includegraphics[width=0.65\linewidth]{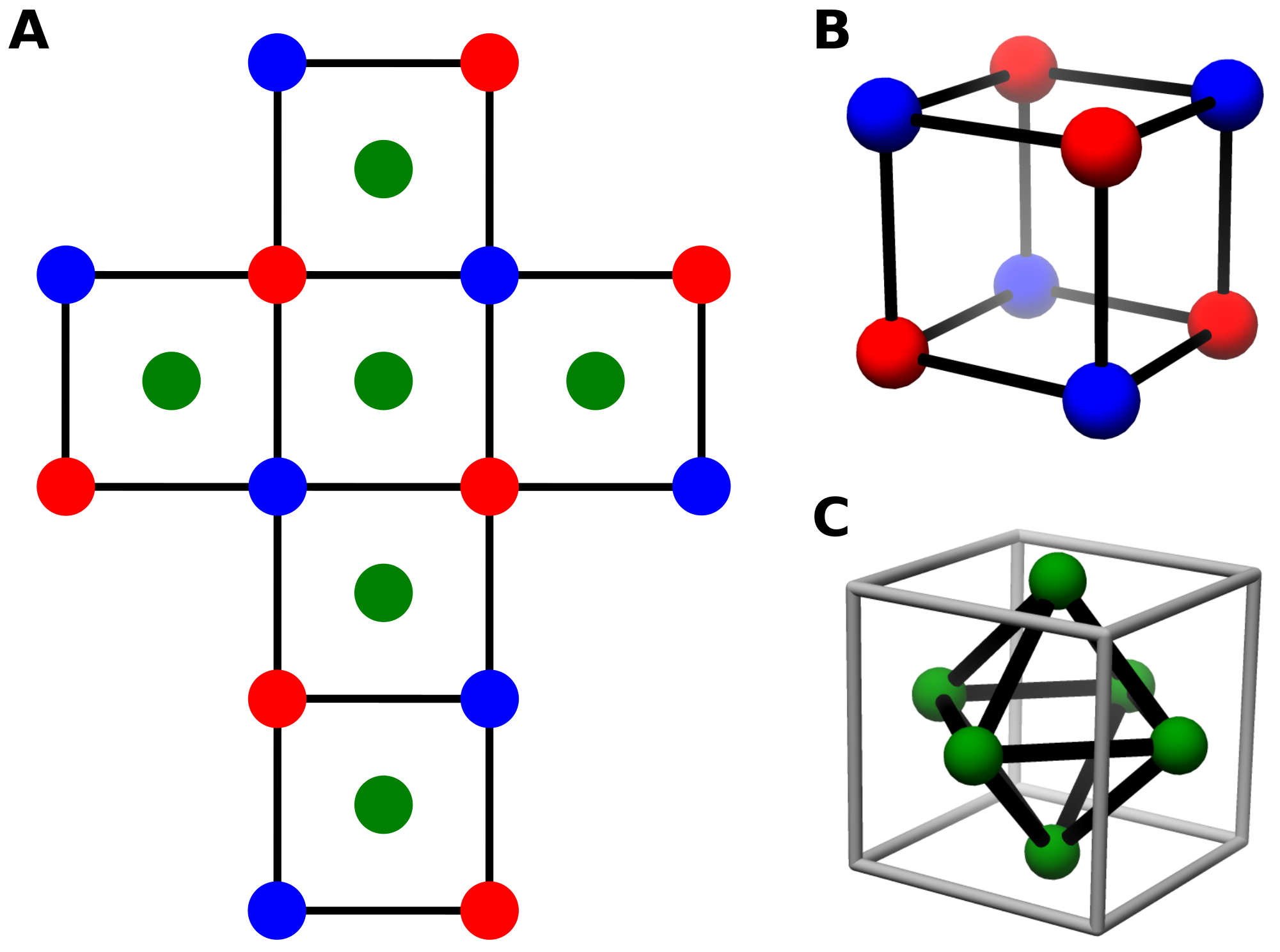}
	\caption{{Deriving tetrahedral and octahedral symmetries from a cubic net. (A) A planar embedding of the surface of a cube. If all vertices are identical, the cube has octahedral symmetry, but colouring its vertices in red and blue to match the cube in (B) reduces it to tetrahedral symmetry. (C) As the octahedron is the dual of the cube, both have the same symmetry. The vertices of the octahedron (green) correspond to the 4-fold symmetry axes of the cube.}}
	\label{SI3}
\end{figure}
\newpage
\begin{figure}[H]
	\centering
	\includegraphics[width=0.7\linewidth]{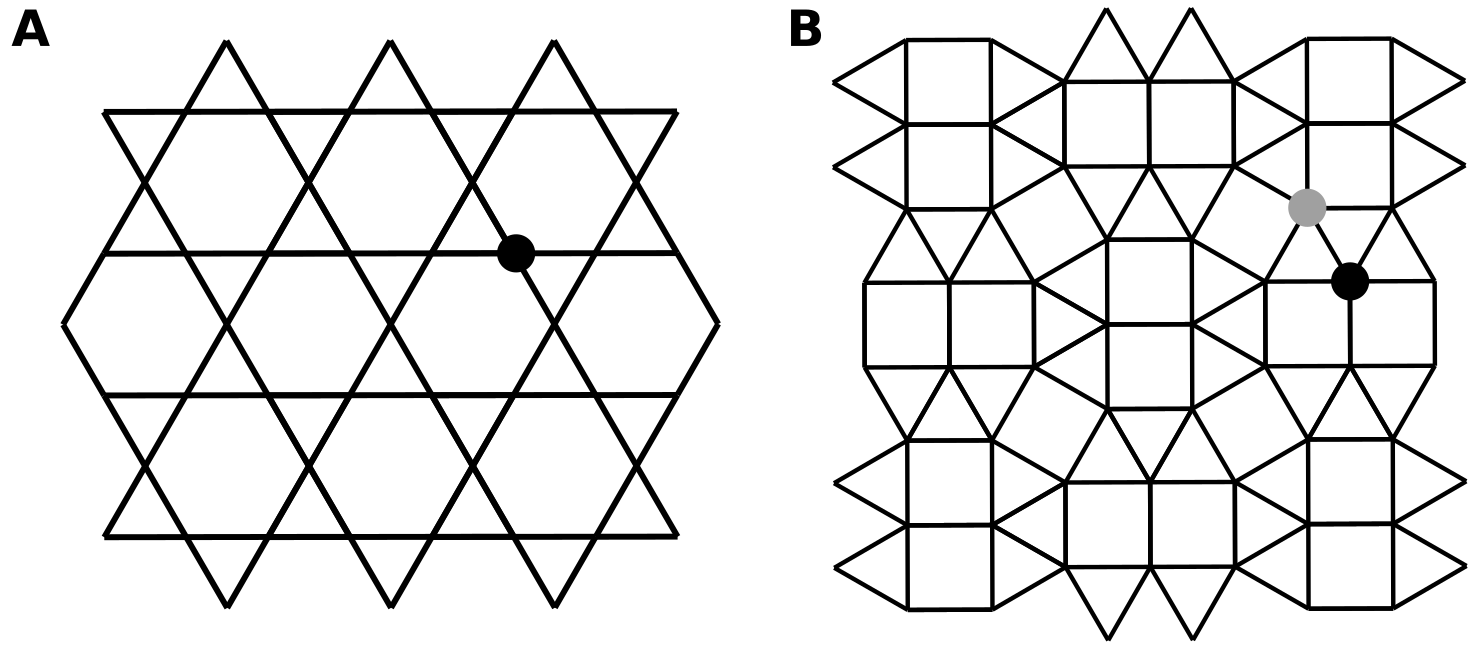}
	\caption{{Examples of a uniform and a $k$-uniform tiling. (A) A uniform tiling is vertex-transitive, i.e. all its vertex environments are identical, as indicated here by a black disk in the ($3\cdot6\cdot3\cdot6$) tiling. (B) In a 2-uniform tiling, each  vertex adopts of one of only two distinct types. In the ($3^3\cdot 4^2$, $3^2\cdot 4\cdot 3\cdot 4$) tiling shown, ($3^3\cdot 4^2$) is indicated by a black disk, and ($3^2\cdot 4\cdot 3\cdot 4$) by a grey disk.}}
	\label{SI4}
\end{figure}
\newpage
\begin{figure}[H]
	\centering
	\includegraphics[width=1\linewidth]{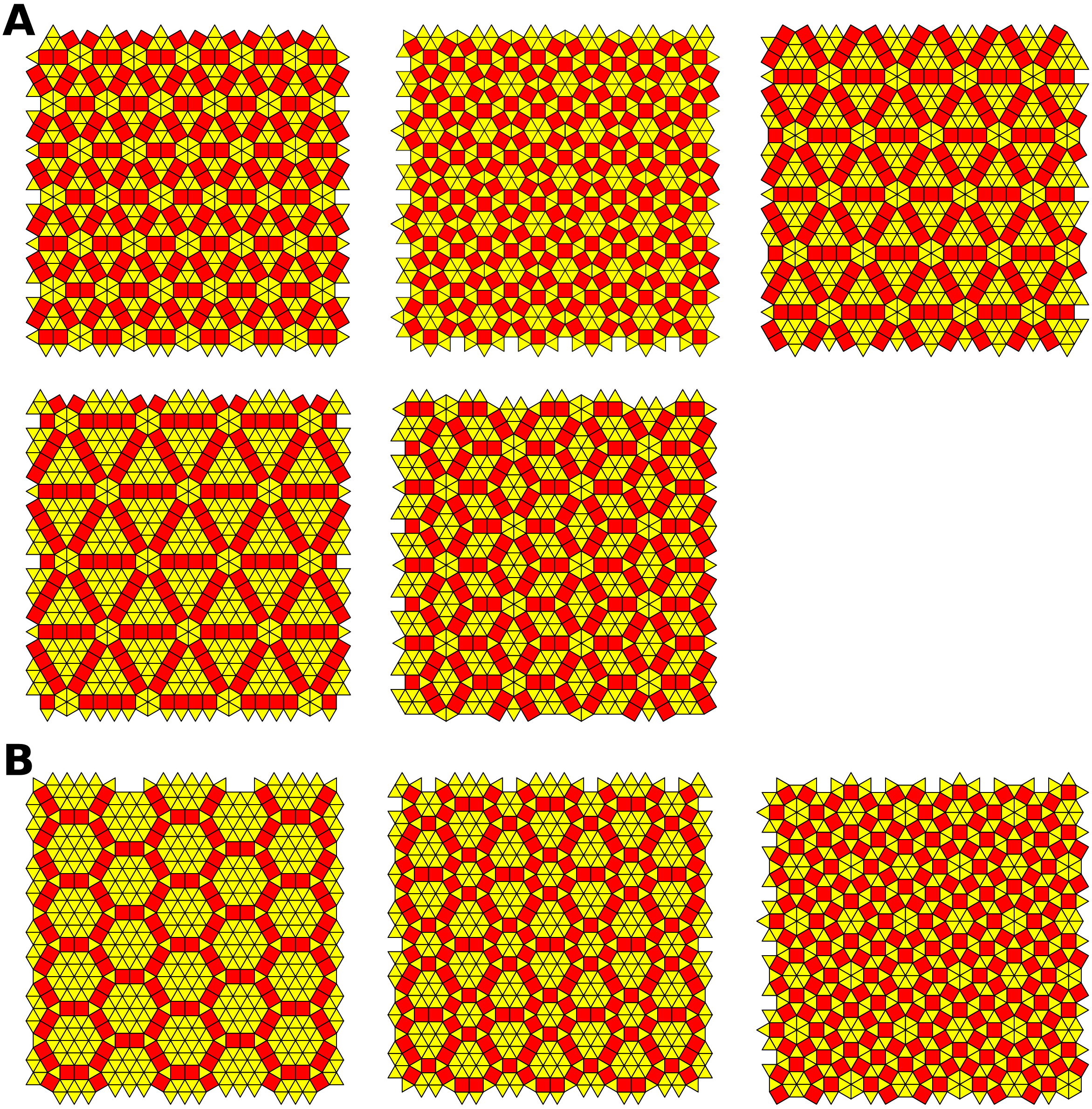}
	\caption{Examples of uniform tilings that do not correspond to surface lattices of AaLS cages. (A) Tilings that cannot be reorganised  into triangles and squashed hexagons do not reflect the local interaction pattern; and (B) tilings that generate spherical particles containing a local 6-fold symmetry axis are incompatible with vertices representing pentamers. Tiling figures are adapted from work by Tom Ruen (\href{https://commons.wikimedia.org/wiki/User:Tomruen}{https://commons.wikimedia.org/wiki/User:Tomruen}) \cite{Tom}.}
	\label{SI5}
\end{figure}
\newpage
\begin{figure}[H]
	\centering
	\includegraphics[width=0.8\linewidth]{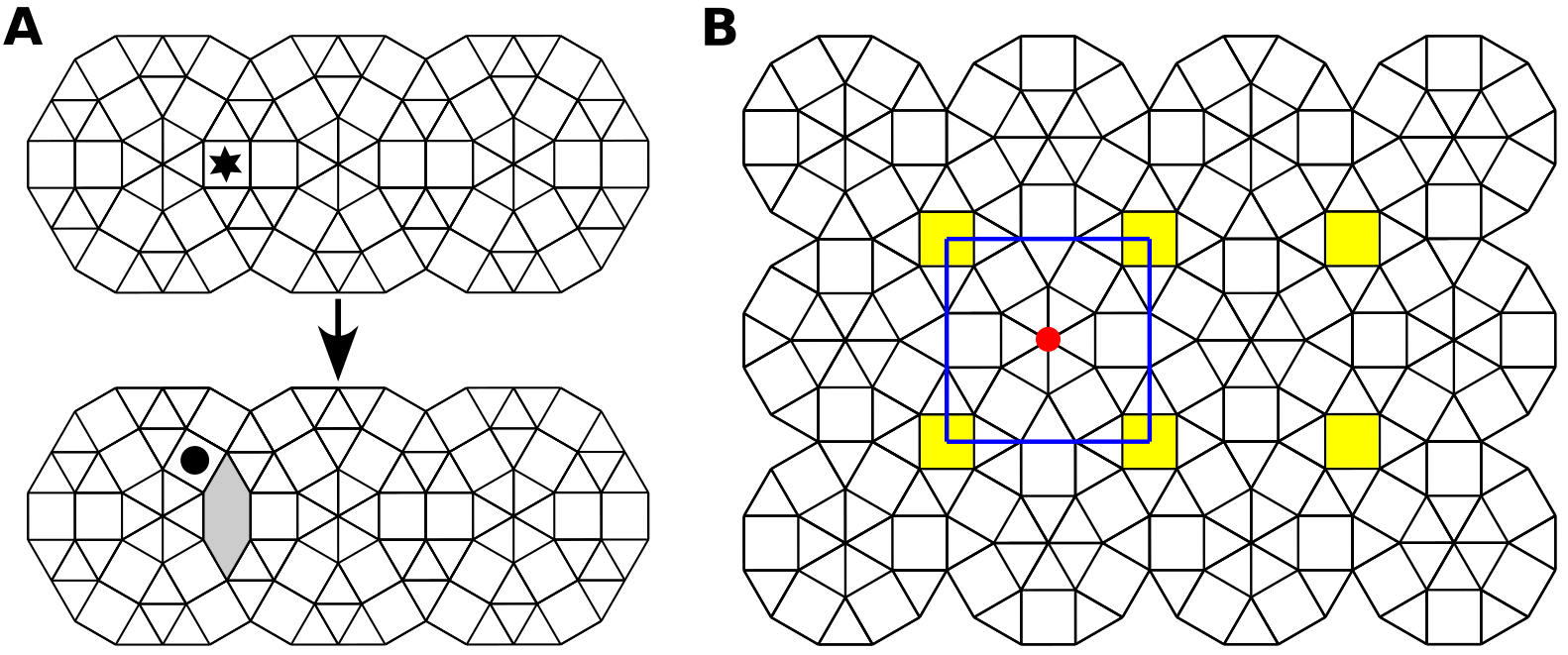}
	\caption{Graphical illustrations of the arguments used to exclude specific tiling types from the classification of AaLS cage architectures. (A) Partitioning the 3-uniform tiling ($3^6$, $3^3\cdot4^2$, $3^2\cdot4\cdot3\cdot4$) into triangles and squashed hexagons is not possible as the centre of the square marked by a disk cannot be the location of a 4-fold symmetry axis, nor can a squashed hexagon be placed on it. (D) The smallest cubic surface that can be embedded into the 5-uniform tiling ($3^6$, ($3^2\cdot4\cdot3\cdot4$)) contains a vertex on a local 6-fold symmetry axis (red dot). As vertices indicate pentamer positions, it is not possible to construct an AaLS cage from this tiling.}
	\label{SI6}
\end{figure}
\newpage
\begin{figure}[H]
	\centering
	\includegraphics[width=0.8\linewidth]{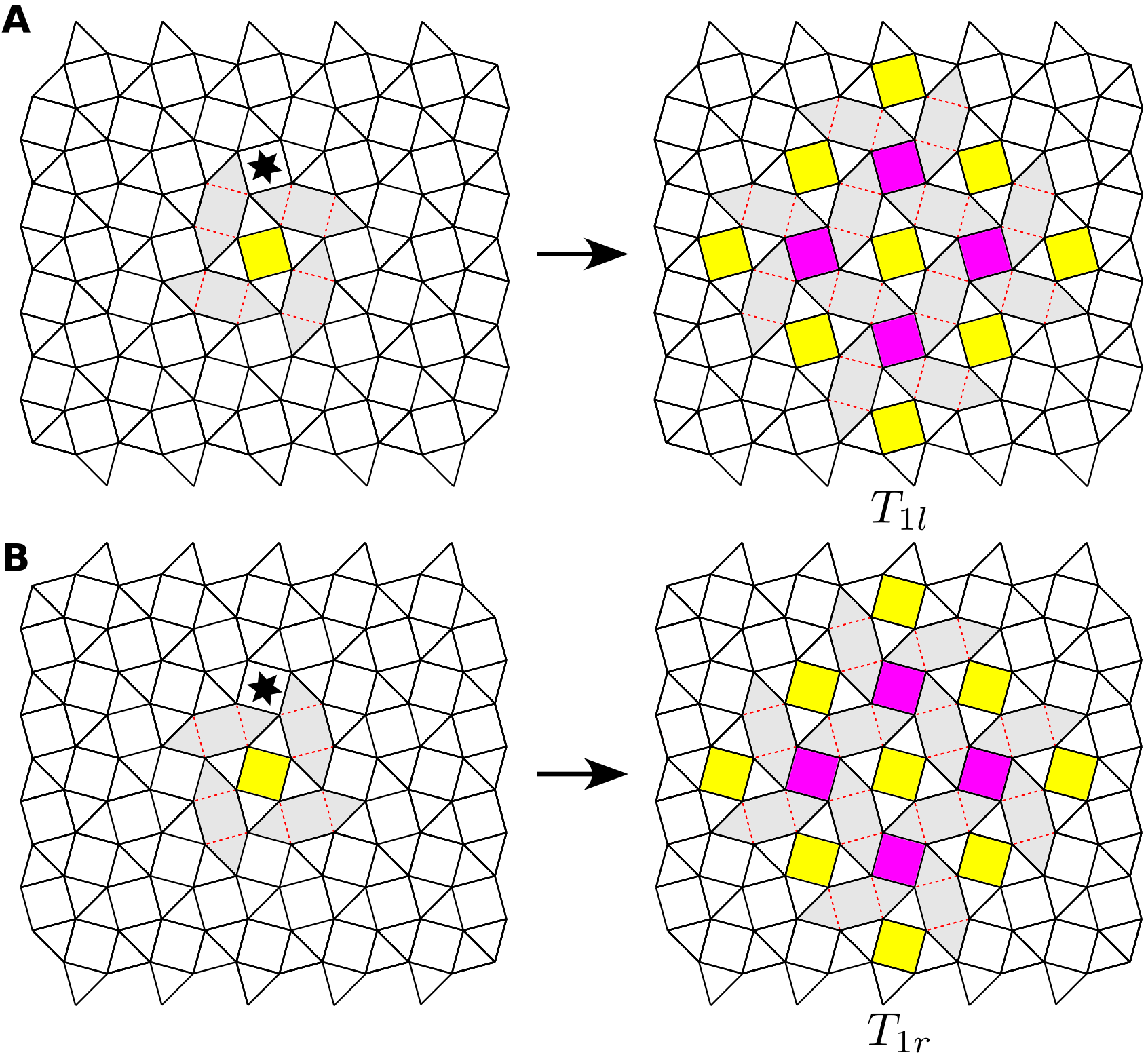}
	\caption{AaLS surface models of opposite handedness. For each tiling in our classification there is a right-handed counterpart that is its mirror image in the plane. For example, the $T_{1l}$ and $T_{1r}$ tilings implied by the arrangements of triangles and squashed hexagons around a left-handed (A) and right-handed (B) \textit{type-1} symmetry axis are mirror images of each other. The squares marked by stars must correspond to left-handed (A) or right-handed (B) \textit{type-2} squares.}
	\label{SI7}
\end{figure}
\newpage
\begin{figure}[H]
	\centering
	\includegraphics[width=0.8\linewidth]{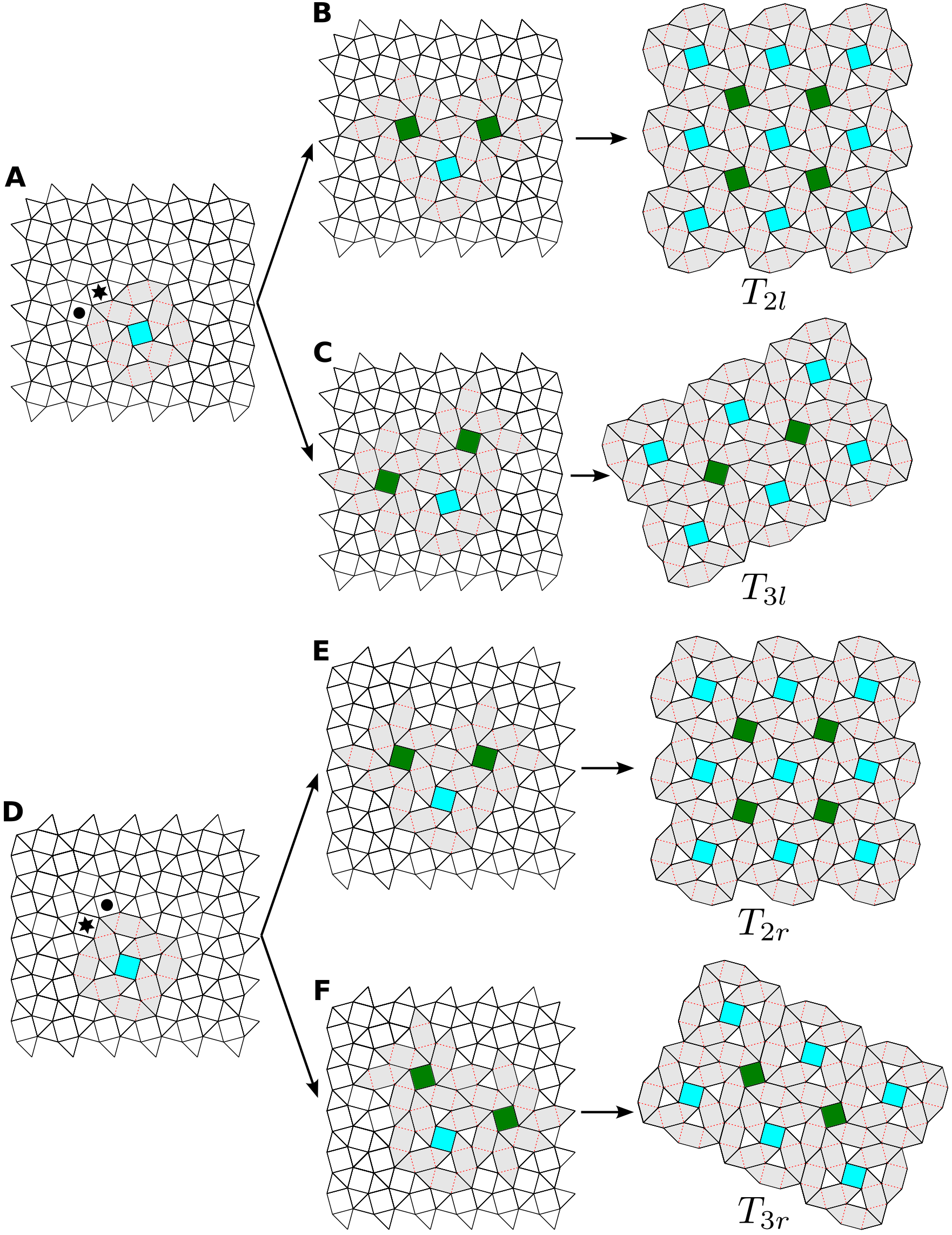}
	\caption{Derivation of AaLS surface models corresponding to \textit{type-3}  and \textit{type-4} squares of opposite handedness. Starting from a left-handed (A) and right-handed (D) \textit{type-3} 4-fold symmetry axis, either the square indicated by a disc or by a star must be a \textit{type-4} square, resulting in the left-handed (B and C) and right-handed (E and F) options, respectively. Each of these can only be continued in a unique way, resulting in left-handed tilings $T_{2l}$ and $T_{3l}$, and right-handed counterparts $T_{2r}$ and $T_{3r}$.}
	\label{SI8}
\end{figure}
\newpage
\begin{figure}[H]
	\centering
	\includegraphics[width=0.75\linewidth]{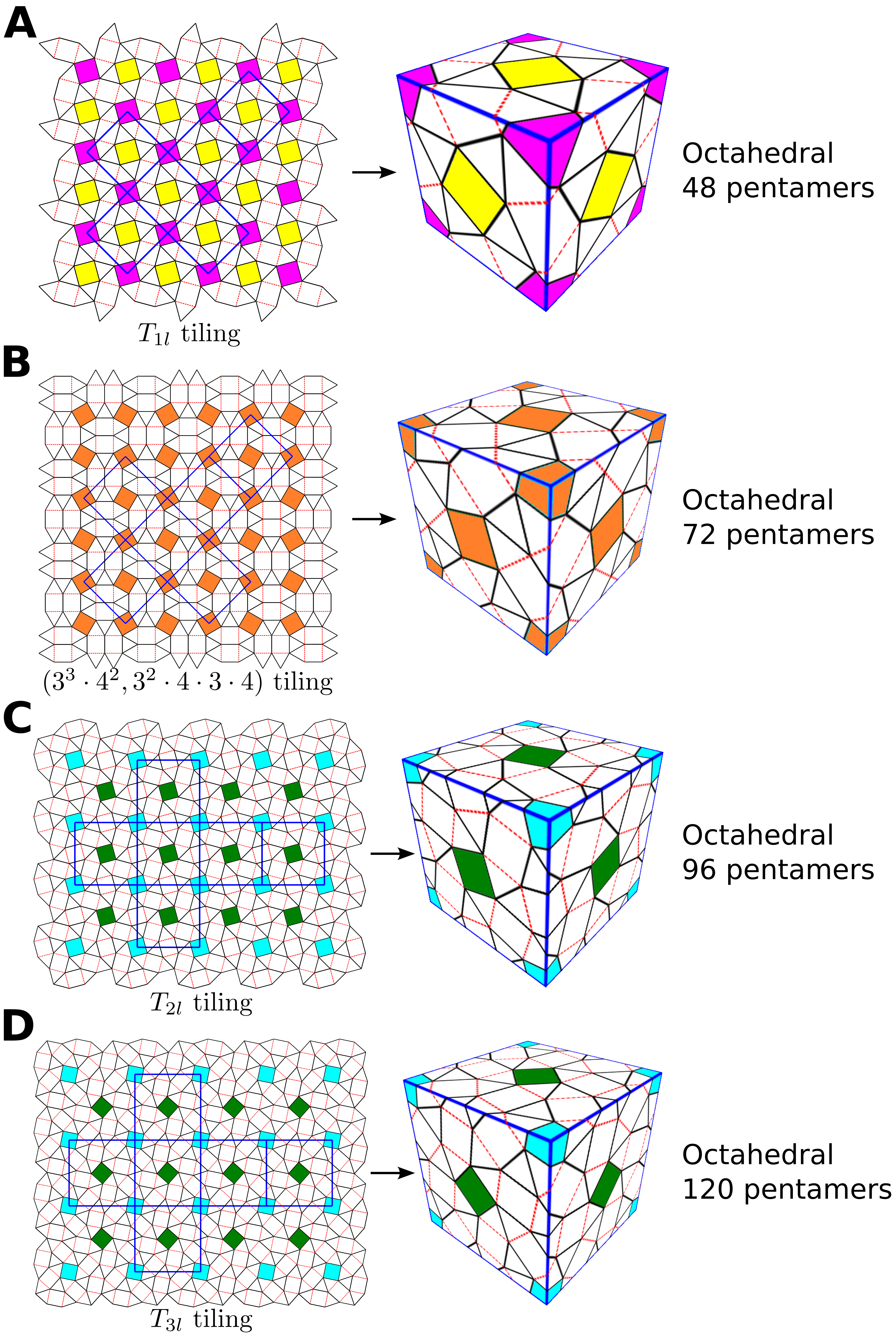}
	\caption{Construction of the second smallest particle layouts from the tilings with 4-fold symmetry axes. Embeddings of rescaled versions of the cubic surface result in octahedral particles that each contain a square in their faces (yellow (A), orange (B), and green (C and D)). As vertices represent pentamers, this would correspond to a hole surrounded by four pentamers interacting with each other in a circular arrangement, which is a local interaction pattern that has not been  observed experimentally. Thus, we reason that these particles are not biologically viable options. However, they might be engineered if pentamers are mutated to enable this type of local interaction.}
	\label{SI9}
\end{figure}
\newpage
\begin{figure}[H]
	\centering
	\includegraphics[width=0.7\linewidth]{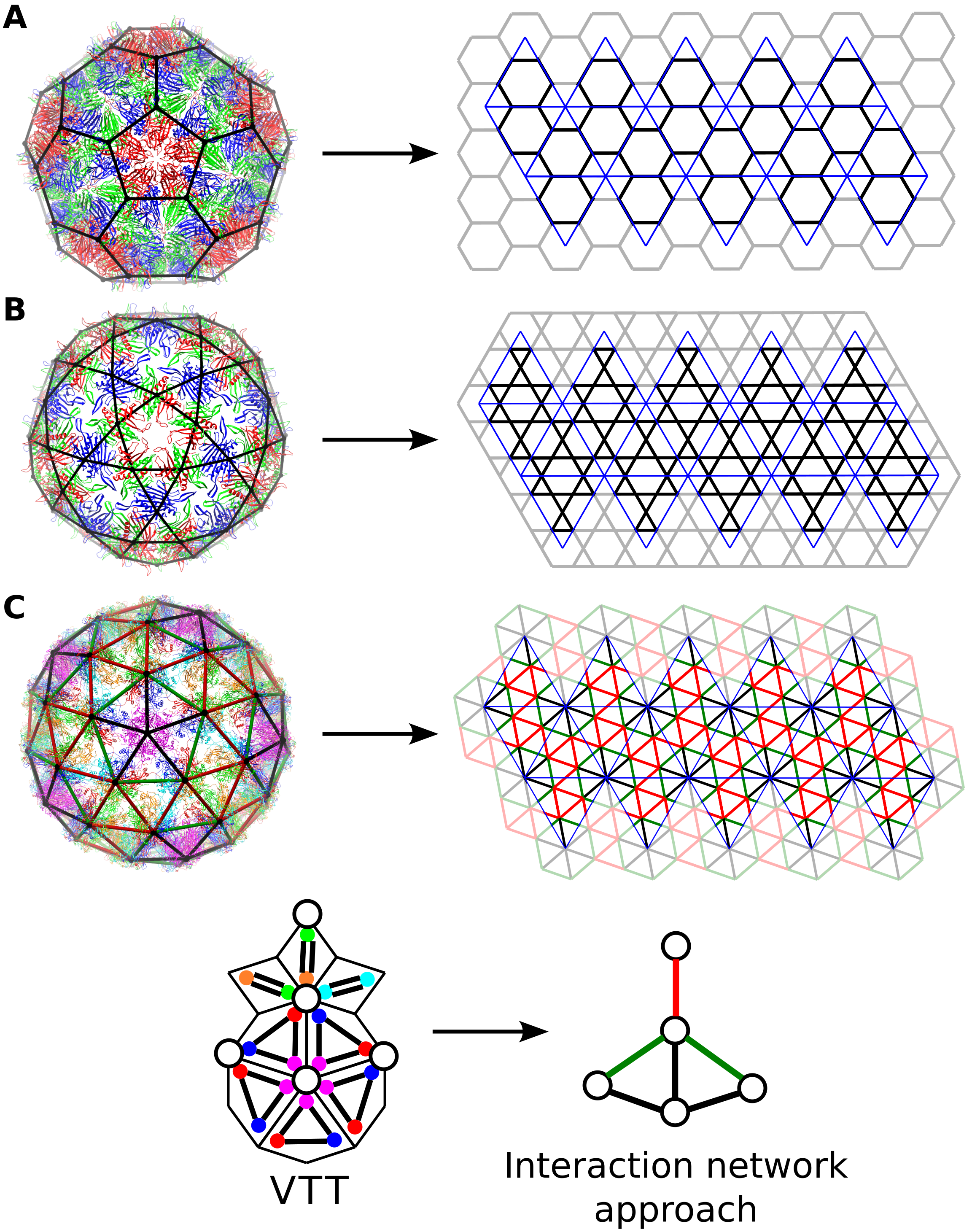}
	\caption{{The tiling models in Caspar-Klug and Viral Tiling theory can also be represented by interaction networks. (A) A hexagonal lattice, connecting the midpoints of the triangular tiles representing the Pariacoto virus capsid (Fig. 1A; PDB: 1F8V), is an example of an interaction network associated with a Caspar-Klug model. (B) The Kagome lattice, consisting of hexagonal and triangular faces, is the interaction network of Bacteriophage MS2 (PDB: 2MS2), which is represented by a rhomb tiling in Viral Tiling theory (VTT). (C) The Human Papilloma virus capsid (PDB: 3J6R), in which protein subunits interact via dimer and trimer interactions, is modelled in terms of rhomb and kite tiles in VTT. Its interaction network is a weighted triangular tiling, in which different types of interactions between pentamers are shown colour-coded. In particular, as the close up at the bottom shows, the dimer and trimer interactions between protein subunits give rise to three distinct interactions between pentamers in the interaction network: red edges correspond to  dimer interactions, i.e. interactions represented by a rhomb tile in VTT; green edges indicate interactions within a trimer, i.e. one kite tile; and black edges represent interactions in two adjacent trimers, i.e. two neighbouring kite tiles.}}
	\label{SI10}
\end{figure}


\end{document}